\documentclass{moriond}

\usepackage{SIunits}

\def\sst{\scriptscriptstyle}

\def\be{\begin{equation}}
\def\ee{\end{equation}}
\def\bea{\begin{eqnarray}}
\def\eea{\end{eqnarray}}

\newcommand{\Photo}{\includegraphics[height=35mm]{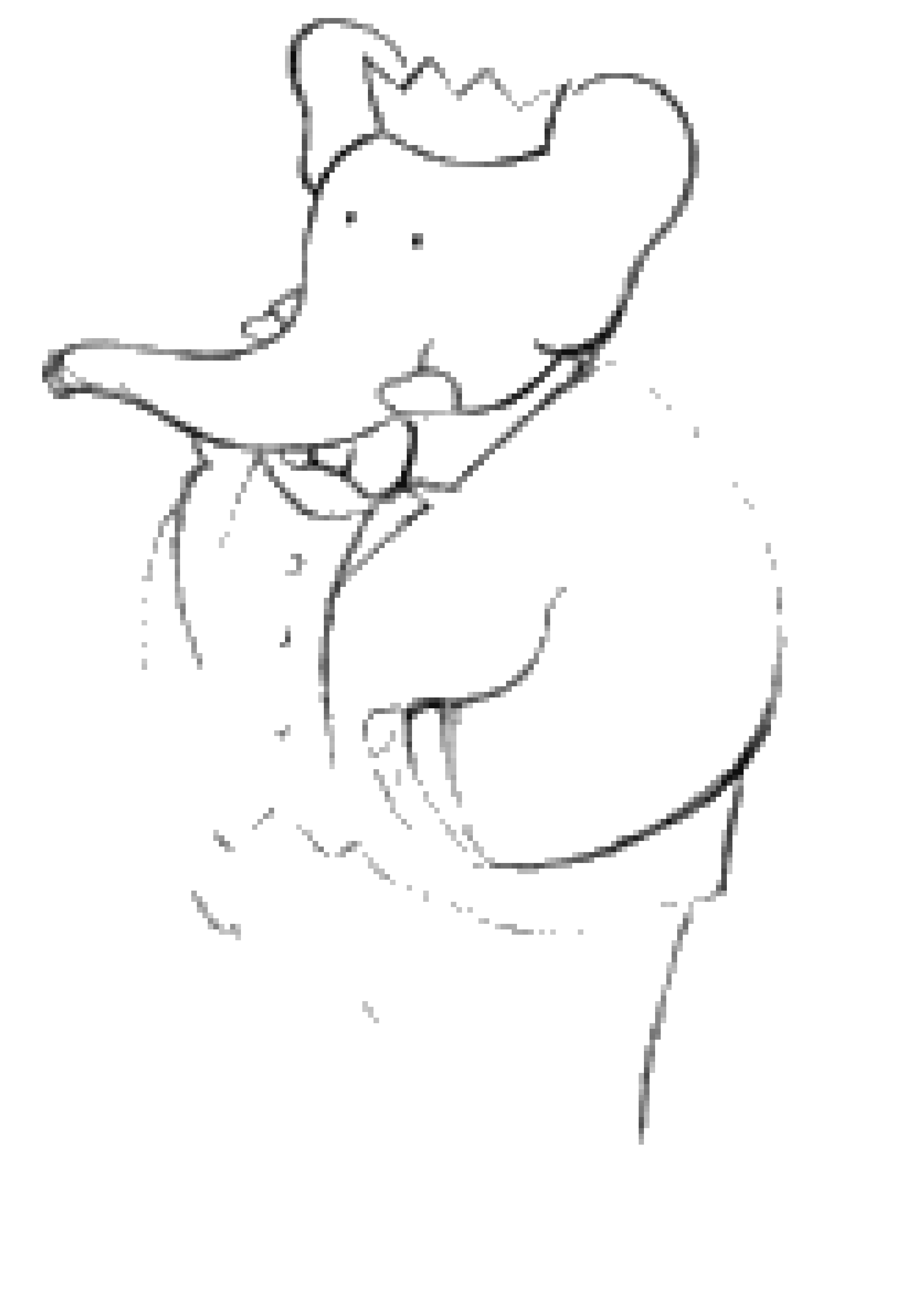}}

\begin{document}
\vspace*{4cm}
\title{Recent BABAR results on measurement of exclusive hadronic cross sections}

\author{D.~Bernard}

\address{Ecole polytechnique, CNRS / IN2P3 and Institut Polytechnique de Paris,
 91128 Palaiseau, France}

\maketitle\abstracts{
The measurement of exclusive $e^+e^-$ to hadrons processes is a significant part of the physics program of $BABAR$ experiment, aimed to improve the calculation of the hadronic contribution to the muon $g-2$ and to study the intermediate dynamics of the processes. We present the most recent results obtained by using the full data set of about 470\,fb$^{-1}$ collected by the BABAR detector at the PEP-II $e^+e^-$ collider at a center-of-mass energy of about 10.6 GeV. In particular, we report the results on $e^+e^-$ annihilation into six- and seven-pion final states. The study of the very rich dynamics of these processes can help to understand the difference seen between the QCD prediction and the sum of exclusive cross sections in the energy region around 2 GeV, thus improving the precision on the total hadronic cross section and of the $g-2$ calculation.
 Additionally, we report the results on a dedicated study to shed light on the resonant states production in the energy region around 2.2 GeV, which is presently rather unclear. We measure the reaction $e^+e^-\to K_SK_L$ with data collected with the $BABAR$ detector and analyse these data in conjunction with published BESIII data on $e^+e^-\to K^+K^-$ and $BABAR$ data on $e^+e^-\to K^+K^-$, $\pi^+\pi^-$, $\pi^+\pi^-\eta$, $\pi^+\pi^-\omega$. This study supports the existence of an isovector resonance $\rho(2230)$ consistent with the resonance observed by BESIII.
}

\begin{center} \bf
On behalf of the BaBar Collaboration
\\
~
\\
 Presented at
 55th Rencontres de Moriond on QCD and High Energy Interactions
\\
 (Moriond QCD 2021)
27 March-3 April 2021.
\end{center}

\section{Introduction}

I report two recent works \cite{BABAR:2019oes,BABAR:2021ywk} by the
BaBar collaboration, both studies of the production of hadronic final
states in $e^+e^-$ collisions with initial state radiation (ISR): one
of the incident leptons first emits a high-energy photon, after which
the $e^+e^-$ collision takes place at a reduced CMS energy.
This enables the measurement of the cross section of the direct
production of that final state in $e^+e^-$ collisions over a large
energy range,\cite{Baier:1969kaa} down to threshold, with excellent
and uniform efficiency, acceptance and background rejection.

\section{Resonances in $e^+e^-$ annihilation near 2.2\,GeV}

We first re-examine a recent publication of the BESIII collaboration
\cite{Ablikim:2018iyx} who compared their result on a resonance
decaying to $K^+K^-$ to a previous result by
BaBar,\cite{Lees:2013gzt,Lees:2015iba} at a mass significantly larger
(Table~\ref{tab:2.2GeV}) than the value quoted for other decay modes
of the $\phi(2170)$ by the PDG
($2160 \pm 80$\,MeV$/c^2$ at the time of writing).
We first perform a fit of the combined BaBar+BESIII sample
\cite{BABAR:2019oes} and obtain a result compatible with the fit of
the BESIII data only \cite{Ablikim:2018iyx} (Table~\ref{tab:2.2GeV}).

\begin{table}[h]
 \caption[]{Results of fits to various final states in the study of
 the resonances in $e^+e^-$ annihilation near
 2.2\,GeV.\cite{BABAR:2019oes} 
 }
\label{tab:2.2GeV}
\vspace{0.4cm}
\begin{center}

\begin{tabular}{|l|l|l|l|l|lllll} \hline
 & $M$ & $\Gamma$ & & \\
 & (MeV/$c^2$) & (MeV) & & \\ \hline
 $K^+K^-$ & $2239\pm7\pm11$ & $140\pm12\pm21$ & BESIII & \cite{Ablikim:2018iyx}
 \\
 $K^+K^-$ & $2227\pm 9\pm 9$ & $127\pm 14\pm 4$ & BaBar+BESIII & \cite{BABAR:2019oes}
 \\
 $I=1$ ($\rho$-like) ($\pi^+\pi^-$ + $\pi^+\pi^- \eta$) & $2270\pm 20\pm 9$ & $116^{+90}_{-60}\pm50$ & BaBar & \cite{BABAR:2019oes}
 \\
$I=0$ ($\omega$-like) ($\omega\pi^+\pi^- + \omega\pi^0\pi^0$)
&
$2265\pm20$ & $75^{+125}_{-27}$ & BaBar & \cite{BABAR:2019oes}
 \\
 \hline
\end{tabular}

\end{center}
\end{table}

If a genuine $q\overline q$ resonance, the decay to $K_SK_L$ should be
visible with the same strength as to $K^+K^-$.
The $K_SK_L$ signal is reconstructed \cite{BABAR:2019oes} from an
ISR-photon candidate with an energy larger than 3\,GeV, a $K_S$ in its
decay to $\pi^+\pi^-$ and a $K_L$ from a cluster in the
electromagnetic calorimeter with an energy larger than 2\,GeV.
Only the direction of flight of the $K_L$ is used in the analysis.
No evidence of a signal is seen but due to limited statistics and the
possible presence of a destructive interference with a non-resonant
contribution, it was not possible to reach a definite conclusion
on an exclusion of a $K_SK_L$ at the same level as for  $K^+K^-$.

As both $I=1$ and $I=0$ amplitudes may contribute to 
$e^+e^-\to KK$
(with different signs for $K^+K^-$ and $K_SK_L$, though), we then turn to
isovector and isoscalar sister channels that BaBar had studied
 (with the ISR method)
in the past:
\begin{itemize} 
\item
 $I=1$:
$\pi^+\pi^-$,(\,\cite{Lees:2012cj}) and 
$\pi^+\pi^- \eta$;(\,\cite{TheBABAR:2018vvb})
\item
 $I=0$:
$\pi^+\pi^-\omega$,(\,\cite{Aubert:2007ef}) and 
$\pi^0\pi^0 \omega$.(\,\cite{Lees:2018dnv})
\end{itemize}

Fits to the combined ($\pi^+\pi^-$ + $\pi^+\pi^- \eta$) samples,
on the one hand,
and to the combined ($\pi^+\pi^-\omega$ + $\pi^0\pi^0 \omega$) samples,
on the other hand,
provided results, again, compatible with the $K^+K^-$ mode
(Table~\ref{tab:2.2GeV}).

\section{Study of
$e^+e^-\to 2(\pi^+\pi^-)\pi^0\pi^0\pi^0$ and $2(\pi^+\pi^-)\pi^0\pi^0\eta$
at CMS energies from threshold to 4.5\,GeV}

All results in this section are first measurements.\cite{BABAR:2021ywk}
The analysis proceeds from the selection of
$ 2(\pi^+\pi^-)\pi^0\pi^0 \gamma \gamma \gamma_{\sst \mathrm{ISR}}$
final states where the four tracks originate from a common vertex
compatible with the interaction region, photon 4-momenta are computed
assuming a direction originating from that vertex and are required to
have an energy $\ge 35$\,MeV, $\pi^0$ are formed from pairs of these
photons, and the highest-energy photon, assumed to be the ISR photon,
$ \gamma_{\sst \mathrm{ISR}}$, has an energy larger than 3\,GeV.
In the case of the channel that has three $\pi^0$ in the final
state, the two candidates having an invariant mass closest to the PDG
value are assigned to $\pi^0$s, and the third one to $\gamma \gamma$.
The selected event sample then undergoes a 6C fit with the 
$ 2(\pi^+\pi^-)\pi^0\pi^0 \gamma \gamma \gamma_{\sst \mathrm{ISR}}$
hypothesis, after which the background ($50 < \chi^2 < 100$) is
subtracted from the signal ($\chi^2 < 50$).

The obtained $\gamma \gamma$ invariant mass spectrum is fit
either with a ($\pi^0 +$ {non-resonant}) pdf, for the low-mass
part of the spectrum,
or with an ($\eta +$ {non-resonant}) pdf, for its high-mass part,
from which the amount of $2(\pi^+\pi^-)\pi^0\pi^0 \pi^0$ and
$2(\pi^+\pi^-)\pi^0\pi^0 \eta $ production can be computed,
and the ISR part is obtained after subtraction of the
$usd$-''continuum'' production.
That fit is also performed on event sub-samples, by bins of the
$2(\pi^+\pi^-)\pi^0\pi^0 \gamma\gamma$ invariant mass, from which the
cross section for
$2(\pi^+\pi^-)\pi^0\pi^0 \pi^0$ and $2(\pi^+\pi^-)\pi^0\pi^0 \eta $
production in $e^+e^-$ collisions as a function of energy is
obtained.

For the $2(\pi^+\pi^-)\pi^0\pi^0 \pi^0$ final state that provides
enough statistics to do so,
(selected with $m_{\gamma\gamma} <0.35\,$MeV$/c^2$),
the fit is also performed on event
sub-samples, by bins of the $3 \pi^0$, $\pi^+\pi^-\pi^0$ or
$\pi^\pm\pi^0$ invariant mass
so as to obtain the
$2(\pi^+\pi^-)\pi^0\pi^0 \gamma\gamma$ invariant mass spectra and the
cross sections of a number of production through
intermediate resonances, including 
\begin{itemize} 
\item
$e^+e^-\to 2(\pi^+\pi^-)\eta$, $\eta\to 3\pi^0$;
\item
$e^+e^-\to\omega\pi^0\eta$, $\omega\to\pi^+\pi^-\pi^0$,
$\eta\to\pi^+\pi^-\pi^0$;
\item
and
$e^+e^-\to\rho(770)^{\pm}\pi^{\mp}\pi^+\pi^-\pi^0\pi^0$,
$\rho(770)^{\pm}\to\pi^{\pm}\pi^0$. 
\end{itemize}

It is interesting to note that the full
$2(\pi^+\pi^-)\pi^0\pi^0 \gamma\gamma$ invariant mass spectrum is very
similar to the spectrum of the sum of all the intermediate-resonance
production, which implies that the contribution of the fully
non-resonant seven-hadron production is a small fraction of the total.

Dedicated analyses of the high $2(\pi^+\pi^-)\pi^0\pi^0 \gamma\gamma$
invariant mass spectra for the full production and for production
through intermediate resonances enable the study of $J/\psi$ and
$\psi(2S)$ decays to these final states.
$B_{J/\psi\to2(\pi^+\pi^-)3\pi^0} = (6.2\pm 0.2\pm 0.9)\times 10^{-2}$
turns out to be the 
largest known branching fraction of the $J/\psi$.\cite{BABAR:2021ywk}

\section{ISR studies at BaBar, status and perspectives}

ISR hadronic production is a key ingredient in the computation of the
hadronic vacuum polarization (HVP) contribution to the standard-model
prediction of the so-called anomalous magnetic moment of the muon
$a_\mu \equiv (g_\mu - 2)/2$.

\begin{table*}[!ht]
\caption{Summary of the BaBar results on ISR production of exclusive
 hadronic final states
(The superseded results have been removed). 
 LO: leading order;
 NLO: next-to-leading order (including one additional photon).
\label{tab:compilation}
}

\begin{center} 
 \small 
\begin{tabular}{|l|c|l|l|llllll}
\hline \noalign{\vskip2pt}
Channels & $\int {\cal L} \mbox{d} t$ (fb$^{-1}$) & Method & Ref. \\
 \noalign{\vskip1pt}
 \hline \noalign{\vskip3pt}
 $2(\pi^+\pi^-)\pi^0\pi^0\pi^0$, 
 $2(\pi^+\pi^-)\pi^0\pi^0\eta$ & 469 & LO & \cite{BABAR:2021ywk}
\\
 $\pi^+\pi^-\pi^0\pi^0\pi^0$, $\pi^+\pi^-\pi^0\pi^0\eta$ & 469 & LO & \cite{Lees:2018dnv}
 \\
 $\pi^+\pi^-\pi^0\pi^0$ & 454 & LO & \cite{TheBaBar:2017vzo}
\\
 $\pi^+\pi^-\eta$ & 469 & LO & \cite{TheBABAR:2018vvb}
\\
$K_SK_L\pi^0$, $K_SK_L\eta$, and $K_SK_L\pi^0\pi^0$ & 469 & LO & \cite{TheBABAR:2017vgl}
\\
$K_SK^+\pi^- \pi^0$, $K_SK^+\pi^- \eta$ & 454 & LO & \cite{TheBABAR:2017aph}
\\
$K^+K^-$ & 469 & LO, no $\gamma_{\sst \mathrm{ISR}}$ tag & \cite{Lees:2015iba}
\\
$K_SK_L$, 
$K_SK_L \pi^+\pi^-$, 
$K_SK_S \pi^+\pi^-$, 
$K_SK_S K^+K^- $ & 469 & LO & \cite{Lees:2014xsh}
\\
$\overline p p $ & 454 & LO & \cite{Lees:2013ebn} 
\\
$\overline p p $ & 469 & LO, no $\gamma_{\sst \mathrm{ISR}}$ tag & \cite{Lees:2013uta} 
\\
$K^+ K^-$ & 232 & NLO & \cite{Lees:2013gzt} 
\\
$\pi^+\pi^-$ & 232 & NLO & \cite{Aubert:2009ad} \cite{Lees:2012cj} 
\\
$2(\pi^+\pi^-)$ & 454 & LO & \cite{Lees:2012cr} 
\\
$K^+K^- \pi^+\pi^-$, 
$K^+K^- \pi^0\pi^0$, 
$K^+K^- K^+K^- $ & 454 & LO & \cite{Lees:2011zi} 
\\
$K^+ K^- \eta$, 
$K^+ K^- \pi^0$, 
$K^0 K^+m \pi^+\pi^-$ & 232 & LO & \cite{Aubert:2007ym} 
\\
$2(\pi^+\pi^-)\pi^0$,
$2(\pi^+\pi^-)\eta$,
 $K^+ K^- \pi^+ \pi^- \pi^0$, $K^+ K^- \pi^+ \pi^- \eta$
 & 232 & LO & \cite{Aubert:2007ef} 
\\
$\Lambda \overline \Lambda $,
$\Lambda \Sigma^0$,
$\Sigma^0 \Sigma^0$
& 232 & LO & 
\cite{Aubert:2007uf} 
\\
$3(\pi^+\pi^-)$, 
$2(\pi^+ \pi^- \pi^0)$, 
$K^+ K^- 2(\pi^+\pi^-)$ & 232 & LO & \cite{Aubert:2006jq} 
 \\
$\pi^+ \pi^- \pi^0 $ & 89 & LO & \cite{Aubert:2004kj} 
 \\
\hline
\end{tabular}
\end{center}
\end{table*}

The historical precise experimental measure of $a_\mu$ at BNL
\cite{Bennett:2006fi}
has been known to depart from the prediction, something which might be
an indication of the presence of new physics beyond the standard model
and has triggered a number of experimental and theoretical works.

In particular, as even though the HVP is a quite small part of the
prediction grand total, its uncertainty is actually the dominant
contribution, and as the contribution of low energy QCD processes are
computed from the measured cross sections of hadronic production in
$e^+e^-$ collisions, the BaBar collaboration has conducted a
systematic campaign of ISR production of exclusive hadronic final
states, that is close to completion.
Thanks to all these efforts, and in particular the to improvement in
the precision brought by BaBar, the prediction-to-measurement
discrepancy for $a_\mu$ amounted to a significance of $3.7$
standard deviations up to recently.\cite{Aoyama:2020ynm}
After having analyzed her run-1 data, the new muon $g-2$ experiment at
Fermilab has measured a value of $a_\mu$ that is compatible with that
of the E821 Experiment at BNL, and that confirms a discrepancy with
the standard-model predicted value that now amounts to $4.2$ standard
deviations.\cite{1856627}
Future updates to a larger statistics by the Fermilab experiment and
results by the J-PARC experiment\cite{Abe:2019thb} will be eagerly welcome.

The BaBar collaboration is presently working on updates of the 
$\pi^+\pi^-$ (\,\cite{Aubert:2009ad,Lees:2012cj}) 
and of the 
$\pi^+ \pi^- \pi^0 $ (\,\cite{Aubert:2004kj})
analyses with an increase of the data sample of a factor of $\approx 2$
and $\approx 5$, respectively, and the most recent version of the
BaBar reconstruction software.
These two channels have a large cross section down to low invariant
masses, something which is of particular importance for the prediction
of the HVP contribution to $a_\mu$ due to the 
$1/s^2$ factor in the dispersion integral
(See, e.g., eq. (109) of the review\cite{Jegerlehner:2009ry}).

\section*{Acknowledgments}

I am immensely grateful to the BaBar collaboration for having invited
me to present these works at Moriond QCD 2021.


\end{document}